\documentclass[aip,apl,reprint,superscriptaddress]{revtex4-1}

\usepackage{graphicx}
\usepackage{dcolumn}
\usepackage{bm}
\usepackage{amsmath}

\begin{document}

\title{Accurate Ab-initio Predictions of III-V Direct-Indirect Band Gap Crossovers}


\author{Jeremy W.~Nicklas}
\affiliation{Department of Physics, The Ohio State University, Columbus, Ohio
43210, USA}

\author{John W.~Wilkins}
\affiliation{Department of Physics, The Ohio State University, Columbus, Ohio
43210, USA}


\date{\today}

\begin{abstract}
We report the compositional dependence of the electronic band structure for a range of III-V alloys. Density functional theory with the PBE functional is insufficient to mimic the electronic gap energies at different symmetry points of the Brillouin zone. The HSE hybrid functional with screened exchange accurately reproduces the experimental band gaps and, more importantly, the alloy concentration of the direct-indirect gap crossovers for the III-V alloys studied here: AlGaAs, InAlAs, AlInP, InGaP, and GaAsP.
\end{abstract}


\maketitle


Knowing the alloy concentrations where the semiconductor is direct or indirect is essential for optoelectronic device design. Consequently a correct quantitative assessment of the electronic structure across different symmetry points of the Brillouin zone for the alloy is vital. Most studies focused on density functional theory (DFT) calculations using the local density approximation (LDA) or the generalized gradient approximation (GGA) which do poorly on excited states. Only recently have hybrid functionals been used in predicting accurate excited states in individual semiconducting alloys.~\cite{Lee2006,Moses2010}

The Heyd-Scuseria-Ernzerhof (HSE) hybrid functional~\cite{Heyd2003,Heyd2006,Krukau2006,Heyd2004,Heyd2005}, which combines the screened exchange with the Perdew-Burke-Ernzerhof (PBE) GGA functional,~\cite{Perdew1996} out performs previous DFT methods in reproducing bulk band gaps.~\cite{Heyd2004,Paier2006-1,Paier2006-2} We report HSE reproduces not only the band gaps across the entire composition range of each alloy studied here but also the direct-indirect band gap crossovers seen experimentally.

\begin{figure}
  \centering
  \includegraphics[]{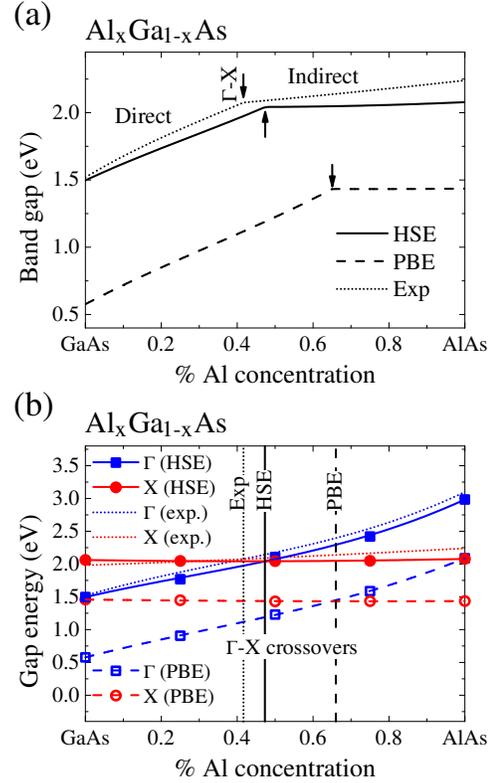}
  \caption{\label{fig:algaas}(Color online) (a) Band gap energy for Al$_x$Ga$_{1-x}$As calculated using HSE (solid line) and PBE (dashed line) in comparison with Vurgaftman \textit{et al.}~\cite{Vurgaftman2001} recommended experimental fit (dotted line). A direct-indirect band gap crossover (vertical arrows) occurs as the Al concentration is increased. (b) A detailed plot of the individual direct (blue), $\Gamma$, and indirect (red), X, gap energy fits using the HSE (solid symbols) and PBE (open symbols) values. The vertical lines denote the direct-indirect band gap crossover.}
\end{figure}

Figure~\ref{fig:algaas}(a) demonstrates this significant improvement of HSE over PBE in predicting the direct-indirect ($\Gamma$-X) crossover (denoted by vertical arrows) for the AlGaAs alloy compared with experiment.~\cite{Vurgaftman2001} HSE reproduces the direct-indirect crossover within 5\% Al concentration from the most recent experimental value published by Yi \textit{et al.}~\cite{Yi2009} The PBE functional which doesn't take into account screened exchange overestimates this crossover by 23\% Al concentration. 

The disordered zinc-blende (cubic) alloys are best modeled by special quasirandom structures (SQS),~\cite{Wei1990} ordered structures designed to reproduce the most important pair-correlation functions of a random alloy. The best possible 32-atom SQS's we produce for concentrations of $x=0.25$ and $x=0.50$ match the pair-correlation functions of a random alloy up to 3rd and 7th nearest neighbors, respectively.~\cite{Nicklas2010} The SQS with concentration $x=0.25$ can be used interchangeably with that of $x=0.75$.

\begin{table}
\caption{\label{tab:supercell}The lattice vectors of the triclinic $x=0.25$ and the base-centered orthorhombic $x=0.50$ SQS's in units of $a/2$ where $a$ is the lattice constant of the alloy A$_x$B$_{1-x}$C.}
\begin{ruledtabular}
\begin{tabular}{lccc}
Concentration & \multicolumn{3}{c}{Lattice vectors} \\
\hline
$x=0.25$ & (1,1,-2) & (0,-3,-1) & (-4,1,-1) \\
$x=0.50$ & (4,2,2) & (-4,2,2) & (0,-1,1)\\
\end{tabular}
\end{ruledtabular}
\end{table}

Table~\ref{tab:supercell} gives the lattice vectors for each SQS used to describe the optical transitions as seen in the zinc-blende primitive cell through folding relations in the Brillouin zone. The SQS with concentration $x=0.50$ has a base-centered orthorhombic space group symmetry with the following folding relations in reciprocal coordinates,
\begin{eqnarray*}
\Gamma^{ZB}(0,0,0) &\rightarrow& \overline{\Gamma}(0,0,0)\\
\textrm{X}^{ZB}(\tfrac{1}{2},\tfrac{1}{2},0) &\rightarrow& \overline{\Gamma}(0,0,0),\;2 \times \overline{\textrm{Z}}(0,0,\tfrac{1}{2})\\
\textrm{L}^{ZB}(\tfrac{1}{2},\tfrac{1}{2},\tfrac{1}{2}) &\rightarrow& 2\times\overline{\Gamma}(0,0,0),\;2 \times\overline{\textrm{Z}}(0,0,\tfrac{1}{2})
\end{eqnarray*}
where the bar denotes superlattice states and the coefficent denotes degeneracies. The SQS with concentration $x=0.25$ has a triclinic space group symmetry with folding relations given by,
\begin{eqnarray*}
\Gamma^{ZB}(0,0,0) &\rightarrow& \overline{\Gamma}(0,0,0)\\
\textrm{X}^{ZB}(\tfrac{1}{2},\tfrac{1}{2},0) &\rightarrow& \overline{\textrm{X}}(\tfrac{1}{2},0,0),\; \overline{\textrm{T}}(0,\tfrac{1}{2},\tfrac{1}{2}),\; \overline{\textrm{R}}(\tfrac{1}{2},\tfrac{1}{2},\tfrac{1}{2})\\
\textrm{L}^{ZB}(\tfrac{1}{2},\tfrac{1}{2},\tfrac{1}{2}) &\rightarrow& \overline{\Gamma}(0,0,0),\; \overline{\textrm{X}}(\tfrac{1}{2},0,0),\; \overline{\textrm{T}}(0,\tfrac{1}{2},\tfrac{1}{2}),\\
&& \overline{\textrm{R}}(\tfrac{1}{2},\tfrac{1}{2},\tfrac{1}{2})
\end{eqnarray*}

The calculations are performed using the projector augmented-wave (PAW) method.~\cite{Blochl1994} The functionals included are the PBE and the HSE06~\cite{Heyd2006} hybrid functional in the \textsc{vasp} code.~\cite{Kresse1996} The Ga 3\textit{d} and In 4\textit{d} electrons are treated as valence and the wavefunctions are expanded in plane waves up to an energy cutoff of 500 eV. The Brillouin-zone integrations have been carried out on (10$\times$10$\times$10), (6$\times$4$\times$4), and (4$\times$4$\times$8) $\Gamma$-centered $k$ meshes for the face-centered cubic primitive cell and SQS supercells for $x=0.25$ and $x=0.50$, respectively.

The lattice constants are linearly interpolated between the experimental parent compound lattice constants taken from Vurgaftman \textit{et al.}.~\cite{Vurgaftman2001} Relaxations are not taken into account after observing only a slight shift of 1\% Al concentration in the direct-indirect crossover in the AlInP alloy using the PBE functional. The compositional dependence of the band gap is described by a quadratic fit to the data, whereas a cubic fit is taken for only the direct gap of AlGaAs.~\cite{Nicklas2010}

\begin{table}
\caption{\label{tab:crossovers}Direct-indirect crossover points for five important semiconducting alloys obtained using HSE and PBE compared alongside experimental measurements. HSE reproduces accurate crossover points for all the alloys other than Al$_x$In$_{1-x}$P.}
\begin{ruledtabular}
\begin{tabular}{lccc}
& HSE & PBE & Exp.\\
\hline
\multicolumn{4}{l}{Al$_x$Ga$_{1-x}$As} \\[2pt]
$x_{\Gamma-\textrm{X}}$ & 0.47 & 0.65 & 0.38\footnotemark[1], 0.42\footnotemark[2]\\
\\
\multicolumn{4}{l}{Al$_x$In$_{1-x}$As} \\[2pt]
$x_{\Gamma-\textrm{X}}$ & 0.68 & 0.76 & 0.64\footnotemark[1]\\
\\
\multicolumn{4}{l}{Al$_x$In$_{1-x}$P} \\[2pt]
$x_{\Gamma-\textrm{X}}$ & 0.37 & 0.44 & 0.44\footnotemark[5], 0.34\footnotemark[6]\\
\\
\multicolumn{4}{l}{Ga$_x$In$_{1-x}$P} \\[2pt]
$x_{\Gamma-\textrm{L}}$ & 0.72 & 0.86 & 0.67\footnotemark[1],0.746\footnotemark[3]\\
$x_{\Gamma-\textrm{X}}$ & 0.73 & 0.90 & \\
$x_{\textrm{L}-\textrm{X}}$ & 0.75 & 0.96 & 0.77\footnotemark[4]\\
\\
\multicolumn{4}{l}{GaAs$_{1-y}$P$_y$}\\[2pt]
$x_{\Gamma-\textrm{L}}$ & 0.56 & 0.77 & \\
$x_{\Gamma-\textrm{X}}$ & 0.57 & 0.84 & 0.45\footnotemark[1]\\
$x_{\textrm{L}-\textrm{X}}$ & 0.58 & 0.94 & \\
\end{tabular}
\end{ruledtabular}
\footnotetext[1]{From Ref.~\cite{Vurgaftman2001}, references therein.}
\footnotetext[2]{From Ref.~\cite{Yi2009}.}
\footnotetext[3]{From Ref.~\cite{Novak2005}.}
\footnotetext[4]{From Ref.~\cite{Merle1977}.}
\footnotetext[5]{From Ref.~\cite{Onton1970}.}
\footnotetext[6]{From Ref.~\cite{Ishitani1996} for strained Al$_x$In$_{1-x}$P.}
\end{table}

\begin{figure*}
  \centering
  \includegraphics[]{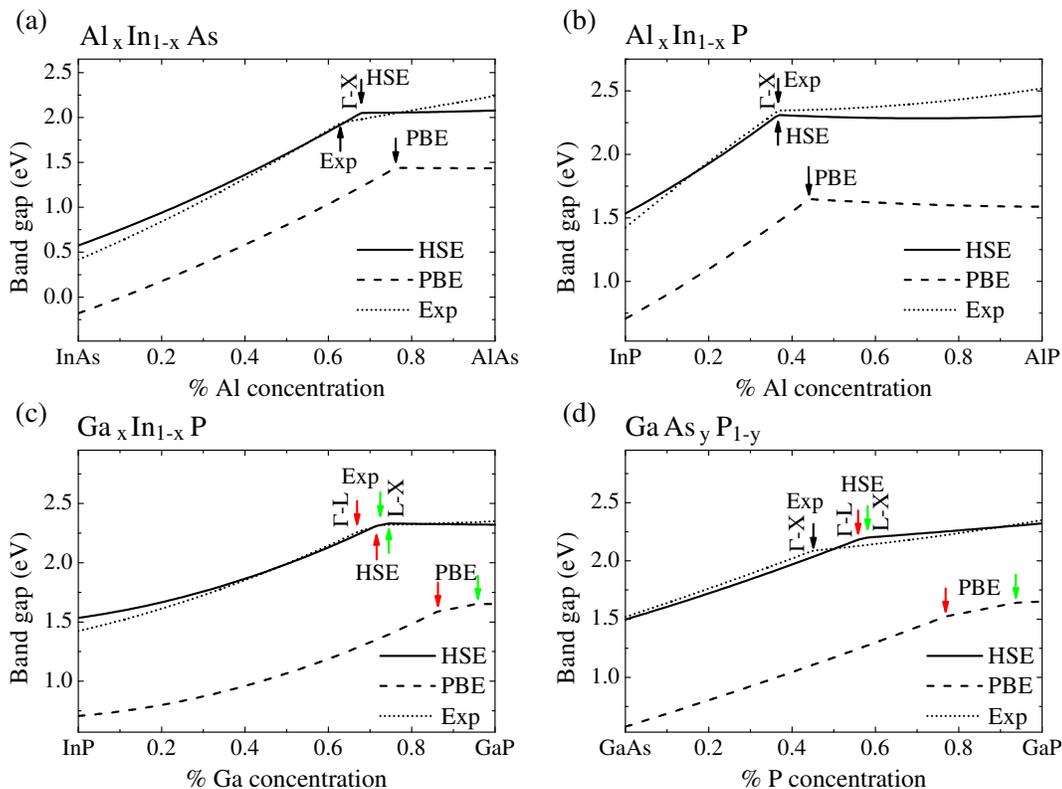}
  \caption{\label{fig:bowing}(Color online) Band gap energy for (a) Al$_x$In$_{1-x}$As, (b) Al$_x$In$_{1-x}$P, (c) Ga$_x$In$_{1-x}$P, and (d) GaAs$_{1-y}$P$_y$ from direct to indirect with increasing concentration. Both HSE (solid line) and PBE (dashed line) are plotted alongside Ref.~\cite{Vurgaftman2001} recommended experimental fits (dotted line). The arrows denote crossovers: $\Gamma$-X (black), $\Gamma$-L (red), and L-X (green). The experimental crossovers here are for visual purposes only and the actual experimental crossovers are listed in Table~\ref{tab:crossovers}.}
\end{figure*}

Figure~\ref{fig:algaas}(b) supplements (a) by detailing the individual direct and indirect gap energies for AlGaAs with the calculated values denoted by symbols. It shows accurate HSE gap energies for both the $\Gamma$ and X symmetry points throughout the entire composition range.

Figure~\ref{fig:bowing} displays the band gap energies and the direct-indirect crossovers for four different III-V alloys. A detailed comparison of the crossovers is given in Table~\ref{tab:crossovers}.

For Al$_x$In$_{1-x}$As, Figure~\ref{fig:bowing}(a) displays the band gap energy for both HSE, PBE, and the recommended experimental bowing parameters.~\cite{Vurgaftman2001} HSE not only predicts a \textit{real} band gap for the concentration range of 0-10\% of Al for this alloy, but also reproduces the $\Gamma$-X crossover by 4\% Al concentration compared with PBE's overestimation by 12\% Al concentration as seen in Table~\ref{tab:crossovers}.

For Ga$_x$In$_{1-x}$P, Figure~\ref{fig:bowing}(c), experiments utilizing optical luminescence measurements see only a single $\Gamma$-X crossover~\cite{Hakki1970,Williams1970,Lettington1971, Chevallier1971,Alibert1972,Macksey1973,Lee1974} whereas high pressure electrical measurements~\cite{Pitt1974} and piezoreflectance measurements~\cite{Merle1977} observed two-point crossovers for $\Gamma$-L and L-X  two point crossovers. Both HSE and PBE yield two point crossovers as seen in Fig.~\ref{fig:bowing}(c) with only HSE reproducing crossover points that lie nearly on top of experiment as well as reproducing the band gap to within 8\% accuracy throughout the whole Ga concentration.

For GaAs$_{1-y}$P$_y$, Figure~\ref{fig:bowing}(d), both HSE and PBE predict a two-point crossover in Fig.~\ref{fig:bowing}(d); whereas experiments which rely on optical luminescence observe only a single point $\Gamma$-X crossover at $x=0.45$. HSE not only produces a direct-indirect crossover that agrees better with experiment, but also predicts a two-point crossover that is only separated by 2\% P concentration that might be seen as a single point crossover in experiment.

For Al$_x$In$_{1-x}$P, Figure~\ref{fig:bowing}(b), HSE predicts a crossover at 37\% Al concentration underestimating the experimental value by Onton \textit{et al.}~($x=0.44$)~\cite{Onton1970} but shows relatively good agreement with the recent results of strained AlInP by Ishitani \textit{et al.}~($x=0.34$).~\cite{Ishitani1996} More studies of the AlInP alloy could provide a detailed comparison with theory.

To conclude, HSE reproduces direct-indirect crossovers within 12\% atomic composition for the alloys studied here, whereas PBE overestimates crossover points by 39\% atomic composition. The HSE functional also substantially improves on band gap energies across the entire composition range. We expect HSE to perform similarly for other semiconducting alloys. 

This work was supported by DOE-Basic Energy
Sciences, Division of Materials Sciences (DE-FG02-99ER45795). This research used computational resources of the National
Energy Research Scientific Computing Center, which is supported by the
Office of Science of the U.S. Department of Energy under Contract No.\
DE-AC02-05CH11231 and the Ohio Supercomputing Center.

\end{document}